\documentclass{moriond}

\usepackage[english]{babel}
\usepackage{amsmath,amssymb}
\usepackage{cleveref}

\usepackage{graphicx}
\usepackage{mdframed}
\usepackage[export]{adjustbox}

\usepackage[dvipsnames]{xcolor}
\usepackage[sort&compress]{natbib}
\usepackage{siunitx}

\bibpunct{[}{]}{,}{n}{}{,}

\def\be{\begin{equation}}
	\def\ee{\end{equation}}
\def\bea{\begin{eqnarray}}
	\def\eea{\end{eqnarray}}

\newcommand{\Photo}{\vspace{3mm}\includegraphics[height=35mm]{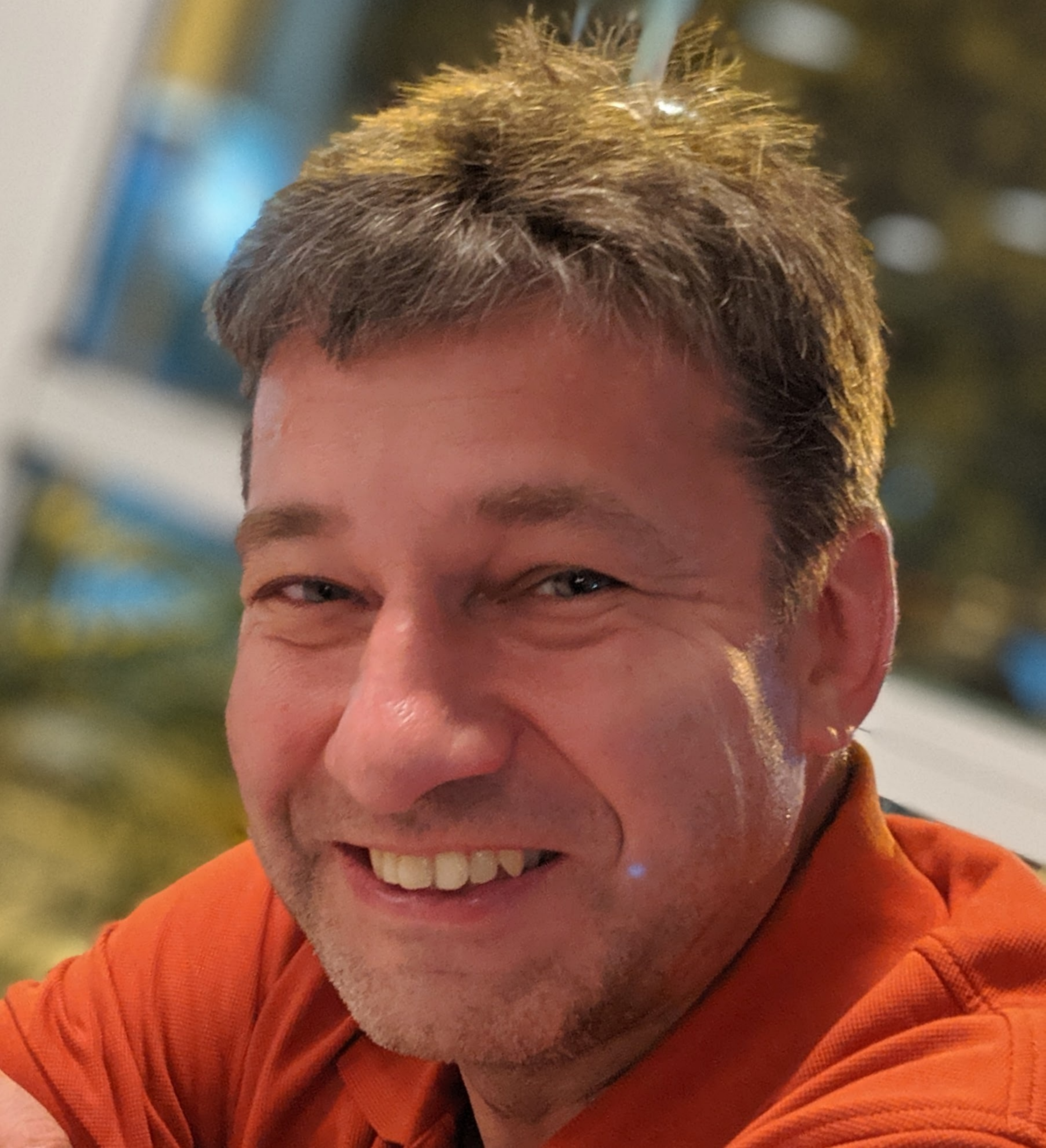}}

\usepackage{diagbox}

\usepackage{booktabs}
\usepackage{subfigure}
\usepackage{multirow}
\usepackage{subfigure}
\usepackage{graphicx}
\usepackage{epsfig}
\usepackage{lscape}
\usepackage{rotating}
\usepackage{epsf}
\usepackage{pstricks}
\usepackage{bm}
\usepackage{slashed}
\usepackage{dsfont}
\usepackage{amsmath}
\usepackage{color}
\usepackage{float}
\usepackage{braket}

\usepackage{stackengine,xcolor}

\usepackage[normalem]{ulem}

\newcommand{\beq}{\begin{eqnarray}}
\newcommand{\eeq}{\end{eqnarray}}

\begin{document}
\vspace*{4cm}
	
\title{Minimal Consistent   models
		for systematic Dark Matter exploration}
	
\author{ \underline{Alexander Belyaev}$^{a,b}$ (speaker),
Giacomo Cacciapaglia$^{c,d}$, Daniel Locke$^{a}$}

\address{
$^a$ School of Physics and Astronomy, University of Southampton, Highfield, Southampton SO17 1BJ, UK
\\
$^b$ Particle Physics Department, Rutherford Appleton Laboratory, Chilton, Didcot, Oxon OX11 0QX, UK
\\
$^c$ Universit\'e de Lyon, F-69622 Lyon, France; Universit\'e Lyon 1, Villeurbanne
\\
$^d$ Institut de Physique Nucl\'eaire de Lyon, CNRS/IN2P3, UMR5822, F-69622 Villeurbanne Cedex, France
}	

\maketitle\abstracts{
Dark Matter search in collider and non-collider experiment requires systematic and consistent approach.
We suggest and perform classification of Minimal Consistent Dark Matter models which aimed to create a solid framework for Dark Matter exploration.
}

\section{Introduction}

Dark Matter (DM) exploration is becoming an increasingly appealing subject at present. Indeed, while
DM evidence from cosmology is one of the strongest  experimental indication of Beyond the Standard Model (BSM) physics, earth-based experiments, including LHC 
do not observe any DM signals.
 At the same time  our knowledge of the nature of DM remains to be unveiled: there are many particle candidates, however no experiment so far was able to probe their properties.

Potentially, DM particles can be probed at the LHC by measuring their production in particle collisions, at direct detection (DD) underground experiments which are sensitive to elastic scattering of DM particles from the local galactic Halo, and in the indirect detection (ID) experiments which measure the product from DM annihilation (and/or decay) in the Universe in the form of positrons, gamma-rays and anti-proton flux.

One of the most important issues behind DM searches is
the lack of a  generic framework which would allow us to combine the results of experimental searches, so different in nature, in a consistent and yet model-independent and general way.

In the last decade the exploration of collider DM phenomenology went beyond the Effective Field Theory (EFT) approach
towards the approach of {\it simplified models}, where the dark matter sector is characterised by the dark matter candidate and a mediator which makes the connection with the SM particles \cite{Buchmueller:2013dya,Cheung:2013dua,Dutta:2014kia,Busoni:2014sya,Papucci:2014iwa,Bai:2014osa,Berlin:2014cfa,Hamaguchi:2014pja,Busoni:2014haa,Balazs:2014jla,Buchmueller:2014yoa,Abdallah:2014hon,Harris:2014hga,Racco:2015dxa,Jacques:2015zha} some of which have been used in recent ATLAS and CMS experimental interpretations.
In case of simplified models the mass of the mediator, and potentially its width, are non-trivial parameters of the model. In these scenarios, one remains agnostic about the theory behind the dark matter sector and tries to parametrise the interactions in the simplest terms: this often leads to writing interactions which are not invariant under the full SM gauge symmetry but only under the unbroken colour SU(3) and electromagnetic U(1). However, the LHC is probing energy scales well above the electroweak symmetry breaking scale, so that for many events the full weak SU(2)$\times$U(1) is a good symmetry.
For instance, if a mediator or DM candidate belongs to a multiplet of the weak SU(2), its charged partners may play an important role in the LHC phenomenology often being more important that the neutral state itself. 
In addition, simplified models often violate gauge invariance which is crucial principle for building the 
consistent BSM model which incorporates the SM together with new physics. For example,
considering simplified model with a new heavy gauge vector boson mediating DM interactions, one should also introduce a
mechanism which is responsible for the mass generation of  this mediator to provide 
gauge invariance for the model which may affect the DM phenomenology. 

The problems and  drawbacks of the previous studies strongly motivate a qualitatively new approach based on building Minimal Consistent Dark Matter (MCDM) models. MCDM models can be still understood as  toy models, that however take in full account the consistency with the symmetries of the SM. Furthermore, a particular MCDM model
can be easily incorporated into a bigger, more complete, BSM model and be explored via complementary constraints 
from collider and direct/indirect DM search experiments as well as relic density constraints
as independent and consistent model.
Another attractive feature of the MCDM approach is their minimal but self-consistent parameter space.

Many implementations of MCDM models are known in the literature~\cite{Deshpande:1977rw,Cirelli:2005uq,Hambye:2009pw,Papucci:2014iwa,Harris:2014hga,Berlin:2014cfa,Bai:2014osa}, however there were no attempt, yet, on their systematic classification. This is precisely the aim of this study, where we   perform a complete classification of MCDM models and  briefly discuss  features of some classes of MCDMs.
Detailed study and discussion will be presented in the follow-up paper~\cite{MCDM-paper}.
\section{Classification of MCDM models}

The building blocks we suggest  to construct models are multiplets defined in terms of their spin and electroweak quantum numbers. We only consider spin-0 ($S$), spin-1/2 ($F$ for a Dirac fermion or $M$ for a Majorana one~\footnote{Here, we consider a Majorana fermion a multiplet with zero $U(1)$ charge and in a real representation of the non-abelian gauge symmetries, such that a mass term $M \bar{\psi}^c \psi$ is allowed.}), and spin-1 ($V$). 
The electroweak quantum numbers will be encoded in the weak Isospin, $I$, and the hypercharge, $Y$, of the multiplet. Furthermore, we will denote with a tilde the multiplets that belong to the dark sector, i.e. they cannot decay into purely SM final states. The multiplets we consider, therefore, read:
$$
\widetilde{S}^I_Y\,, \quad \widetilde{F}^I_Y\,, \quad \widetilde{M}^I_0\,, \quad \widetilde{V}^I_Y\,,
$$
and similarly with un-tilded ones. As some mediator multiplets may carry QCD quantum numbers, we will use a superscript ${}^c$ to label this feature.

To construct consistent minimal models, we follow the main building principles:
\begin{itemize}
	\item[I)] we add one Dark multiplet (including the singlet case) and all its renormalisable interactions with the SM fields, 
	excluding those  that trigger the decays of the multiplet, which is therefore stable. The models will automatically 
	include a Dark symmetry, being $\mathbb{Z}_2$ or $U(1)$ depending on the multiplet. The weak Isospin and 
	hypercharge are constrained by the need of having a neutral component, therefore we will have the following two 
	cases:
	\begin{itemize}
		\item[1)] with integer isospin $I=n$, $n \in N$, so $Y = 0, 1 \dots n$;
		\item[2)] with semi-integer isospin $I = (2n+1)/2$, $n \in N$, so  $Y = 1/2, 3/2 \dots (2n+1)/2$\ \ .
	\end{itemize}
	Note that the case of negative hypercharge can be obtained by considering the charge conjugate field, thus 
	the sign of $Y$ is effectively redundant, and we will consider $Y \geq 0$.

	\item[II)] we consider  models 
		\begin{itemize}
		\item[1)]	
		with just one DM  multiplet and mediators being SM fields.~\footnote{Note that this model building approach has been used in~\cite{Cirelli:2005uq} to construct models of so-called Minimal Dark Matter, so some of the results we present here can be found in this reference. However, our approach has some 
		differences: in Ref.~\cite{Cirelli:2005uq}, the symmetry making the DM candidate stable or long lived emerged as at low energy, at the level of renormalisable interactions, while decays could be induced by higher dimensional couplings to the Higgs multiplets. In our case, we assume that a parity or global $U(1)$ symmetry is also respected by higher dimensional operators.} While our principle is to be limited to renormalisable interactions, under the assumption that higher order ones are suppressed by a large enough scale to make them irrelevant for the DM properties, in some cases we will consider dimension-5 operators.
		\item[2)] we also consider models with just one mediator multiplet, characterised by the respective weak Isospin, $I'$, and hypercharge, $Y'$.	The mediator multiplet can be odd or even with respect to Dark symmetry, and	its quantum numbers are limited to cases where renormalisable couplings to the Dark multiplet and to the SM are allowed. 	This opens the possibility of multiplets carrying QCD charges, which we label with a superscript ${}^c$. The mediators	are labelled as following:
			\begin{itemize}
			\item[a)]  ${S}_{Y'}^{I'(c)}$, ${F}_{Y'}^{I'(c)}$, ${M}_{0}^{I'(c)}$ and ${V}_{Y'}^{I'(c)}$ for even mediator multiplets;
			\item[b)]  $\widetilde{S}_{Y'}^{I'(c)}$, $\widetilde{F}_{Y'}^{I'(c)}$, $\widetilde{M}_{0}^{I'(c)}$ and $\widetilde{V}_{Y'}^{I'(c)}$ for odd mediator multiplets.
		\end{itemize}
	\end{itemize}
	
	\item[III)] we consider all renormalisable interactions allowed by the QFT. Our basic assumption for MCDM models is that higher-order operators are suppressed by a scale high enough that the LHC is unable to resolve the physics generating the operators. The effect on the DM properties is also considered negligible (except for dim-5 operators generating mass splits).
	
	\item[IV)] we ensure cancellation of triangle anomalies, so that the MCDM models entails consistent gauge symmetries, and consider minimal flavour violation (MFV) couplings to SM fermion generations.
\end{itemize}

With the notations above, following the precepts I) to IV), we can classify all MCDM models with up to one mediator multiplet 
using a 2-dimensional Table in Spin(DM)-Spin(mediator) space, as presented in Table~\ref{tab:MCDM}.
Each specific DM model is denoted by a one- or two-symbol notation, indicating the DM multiplet first, followed by the mediator multiplet.
One should note that in this case SM particles as well as members of DM multiplet other than DM, 
could also mediate DM interactions and their interference with the mediator multiplet  can be non-trivial.
Eventually, the case with no mediator multiplet is denoted by just one symbol labelling the DM multiplet.
In this case the role of mediators can only be played by SM particles and members of DM multiplet.

\begin{table}[htb]
	\begin{center}
		\begin{tabular}{||l||c|c|c||}
			%
			\hline
			\diagbox[]{Spin of\\ Mediator}{Spin of \\Dark \\Matter}&
			0&1/2&1
			\\
			\hline
			\hline
			no mediator& 
			$\widetilde{S}_Y^I$
			& 
			$\widetilde{F}_Y^I$
			&
			$\widetilde{V}_Y^I$
			\\
			\hline
			spin 0 even mediator& 
			$\widetilde{S}_Y^I{S}_{Y'}^{I'}$
			& 
			$\widetilde{F}_Y^I S_0^{I'}$
			&
			$\widetilde{V}_Y^I{S}_{Y'}^{I'}$
			\\
			spin 0 odd mediator& 
			$\widetilde{S}_Y^I\widetilde{S}_{Y'}^{I'}$
			& 
			$\widetilde{F}_Y^I \widetilde{S}_{Y'}^{I'}$  \ \ \  $\widetilde{F}_Y^I \widetilde{S}_{Y'}^{I'c}$
			&
			$\widetilde{V}_Y^I\widetilde{S}_{Y'}^{I'}$
			\\
			\hline
			spin $1/2$ even mediator & 
			& 
			
			&
			
			\\&&&\\
			spin $1/2$ odd mediator &
			$\widetilde{S}_Y^I \widetilde{F}_{Y'}^{I'} \ \ \ \widetilde{S}_Y^I \widetilde{F}_{Y'}^{I'c}$
			& 
			$\widetilde{F}_Y^I \widetilde{F}_{Y\pm 1/2}^{I\pm 1/2}$
			&
			$\widetilde{V}_Y^I \widetilde{F}_{Y'}^{I'} \ \ \ \widetilde{V}_Y^I \widetilde{F}_{Y'}^{I'c}$
			\\
			\hline
			spin $1$ even mediator & 
			$\widetilde{S}_Y^I{V}_{0}^{I'}$
			& 
			$\widetilde{F}_Y^I V_0^{I'}$
			&
			$ \widetilde{V}_Y^I {V}_{Y'}^{I'}$
			\\
			spin $1$ odd mediator &
			$\widetilde{S}_Y^I \widetilde{V}_{Y'}^{I'}$
			& 
			$\widetilde{F}_Y^I \widetilde{V}_{Y'}^{I'}$ \ \ \ $\widetilde{F}_Y^I \widetilde{V}_{Y'}^{I'c}$
			&
			$\widetilde{V}_Y^I\widetilde{V}_{Y'}^{I'}$
			\\
			\hline
		\end{tabular}
		\caption{\label{tab:MCDM}Classification of the Minimal Consistent Dark Matter (MCDM) Models
			in Spin(DM)-Spin(mediator) space. When possible, the Dirac fermion can be replaced by a Majorana one, $F \to M$.}
	\end{center}
\end{table}
In the remainder of this paper, we discuss spin-1/2 DM multiplets only, leaving the other two cases for a future publication.


\section{Case of only  DM multiplet: $\tilde{F}^I_Y$ and $\tilde{M}^I_0$ models} \label{sec:DMmult}

Models where the DM belongs to a single EW multiplet, while no other light states are present, have been studied in great detail, starting from the seminal paper in Ref.~\cite{Cirelli:2005uq}. Here we add a detailed discussion of the following novel aspects, including:
a) an improved formula for the mass splitting induced by EW loops, which is numerically more stable than the one given in Ref.~\cite{Cirelli:2005uq};
b) the effect of couplings to the Higgs boson arising from dimension-5 operators. While going beyond the minimality principle, they can be generated by integrating out a single mediator. Furthermore, a class of these operators have special phenomenological relevance as they can help salvage some of the minimal models with non-zero hypercharge; c) up-to date discussion of Direct Detection bounds, including loop-induced interactions.

In the ``stand alone" case, the most general renormalisable Lagrangian for the DM multiplet $\Psi$, with isospin and hypercharge $\{I, Y\}$, is
\beq \label{eq:LagrDirac0}
\mathcal{L} = i \bar{\Psi} \gamma^\mu D_\mu \Psi - m_D \bar{\Psi} \Psi - \frac{1}{2} \left( m_M \; \bar{\Psi}^C \Psi\, + \mbox{h.c.} \right),
\eeq
where the superscript ${}^C$ indicates the charge-conjugate field. We have explicitly added both a Majorana mass $m_M$, which is only allowed for $Y=0$ (thus, integer isospin), and a Dirac one $m_D$, which vanishes for a Majorana multiplet.
This simple class of models has well established properties~\cite{Cirelli:2005uq}.

We should  note that for DM multiplets with $\{I, Y\}  = \{0,0\}$, $\{1/2, 1/2\}$, $\{1,0\}$ and $\{1,1\}$, a linear Yukawa coupling with the SM leptons is allowed by gauge symmetries, while larger isospin multiplets are automatically protected at renormalisable level. However, higher order couplings involving the Higgs can always generate decays of the DM multiplets, and it has been the main motivation of Ref.~\cite{Cirelli:2005uq} to find multiplets that are long-lived enough to be Cosmologically stable. In this work we will be more pragmatic and allow for any multiplet by forbidding implicitly all operators that could mediate the decays of the DM candidate. The origin of such a symmetry is to be searched in the more complete model containing the DM multiplet.

In the case of Dirac multiplets ($\tilde{F}^I_Y$), i.e. when both chiralities are present, the lowest order Lagrangian in Eq.~\eqref{eq:LagrDirac0} is invariant under a global U(1)$_{\rm DM}$ symmetry, thus an asymmetric contribution to the relic abundance may be present if the complete model preserves this symmetry. 
In the case $Y=0$, the presence of a Majorana mass breaks U(1)$_{\rm DM} \to \mathbb{Z}_2$.~\footnote{Note that the Majorana mass is not generated radiatively as long as the U(1)$_{\rm DM}$ symmetry is preserved by the complete model.}

\begin{figure}[tb]
	\begin{center}
		\includegraphics[width=7cm]{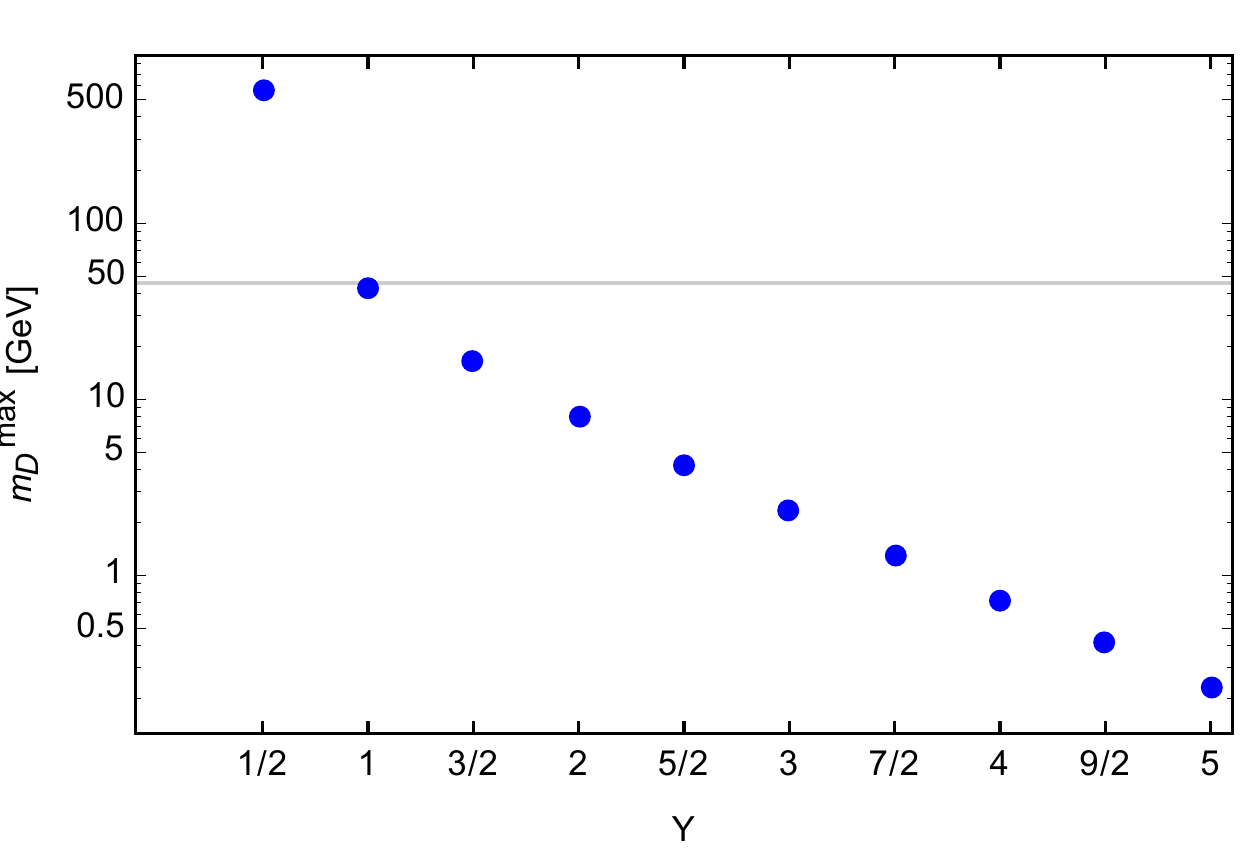} \qquad
		\includegraphics[width=7cm]{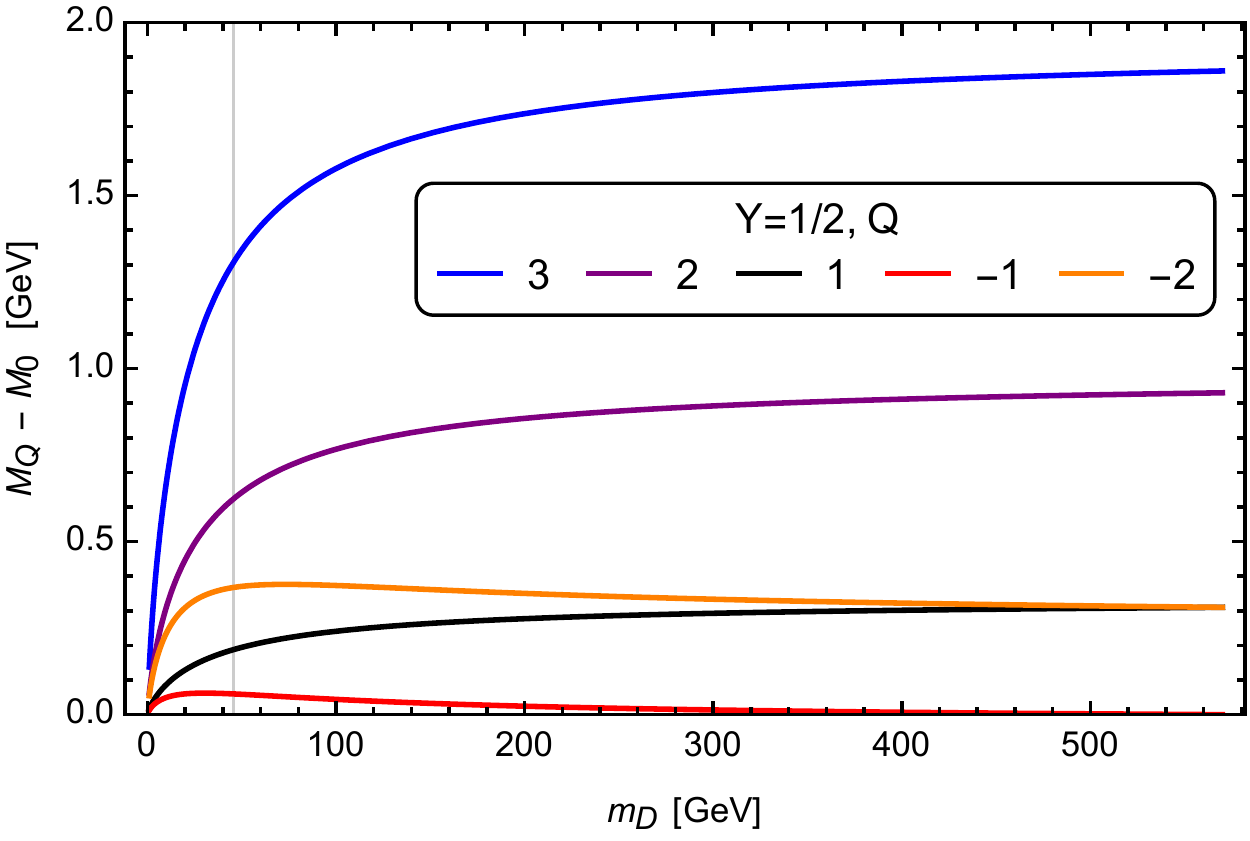}
	\end{center}
\vskip -0.5cm
	\caption{{\it Left}: maximum value of $m_D$ above which the lightest component has charge $Q=-1$ for various values of $Y$. The horizontal line indicates $m_Z/2$, below which decays of the $Z$ exclude the model. {\it Right}: spectrum for a generic multiplet with $Y=1/2$, with $m_D < 570$~GeV. The vertical line shows $m_D \approx m_Z/2$, below which the model is excluded by the $Z$ decays.} \label{fig:deltaM_F}
\end{figure}

Except for the singlet case $\tilde{F}^0_0$, the multiplet contains extra charged states:
\beq
\Psi = \left(
	\psi^{n+} , ... , \	\psi^+ ,
	\psi_0 , \ \psi^- ,\ ... \ \psi^{m-} \right)\,, \qquad \mbox{with} \quad n = I+Y\,, \;\; \mbox{and}\;\; m=I-Y\,. 
\label{eq:Dirac}
\eeq
The Dirac mass term in Eq.~\eqref{eq:LagrDirac0} gives equal mass to all components of the multiplet. This degeneracy is resolved by radiative corrections due to the EW gauge bosons. This contribution has first been computed in Ref.~\cite{Cirelli:2005uq}.  In particular, states with $Q<0$ are always lighter than the $Q=0$ one in this limit.
Thus, there exists an upper limit on $m_D$, above which the lightest state in the multiplet is charged, and this value is determined by the $Q=-1$ state. The values of the mass upper bounds for various $Y$ are shown in the left panel of Fig.~\ref{fig:deltaM_F}: 
the highest value is achieved for $Y=1/2$ which gives $m_D^{\rm max} = 570$~GeV (we recall that for $Y=0$ there is no limit), while for $Y=1$ we find $m_D^{\rm max} = 42$~GeV, which is already below $m_Z/2$. Thus, multiplets with $Y \geq 1$ are excluded by the $Z$-width measurement in the region where the lightest state is neutral. 

\subsection{Direct Detection} \label{sec:DDminimal}
\begin{figure}[htb]\label{fig:triLoop}
\vskip -0.5cm
{\hspace*{-0.5cm}\includegraphics[width=5.5cm]{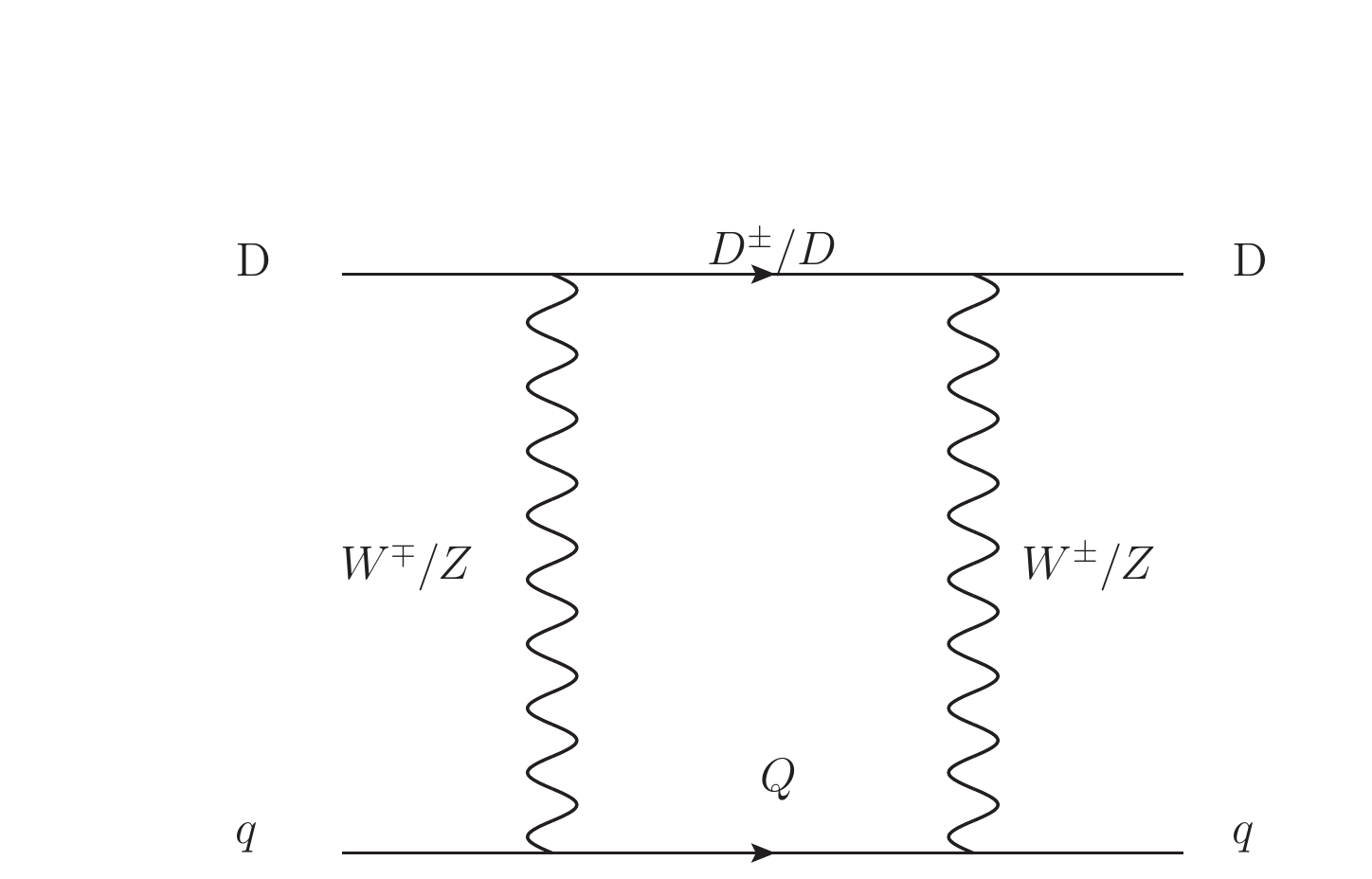}
\\
\vskip -0.5cm
\hspace*{-0.5cm}\includegraphics[width=5.5cm]{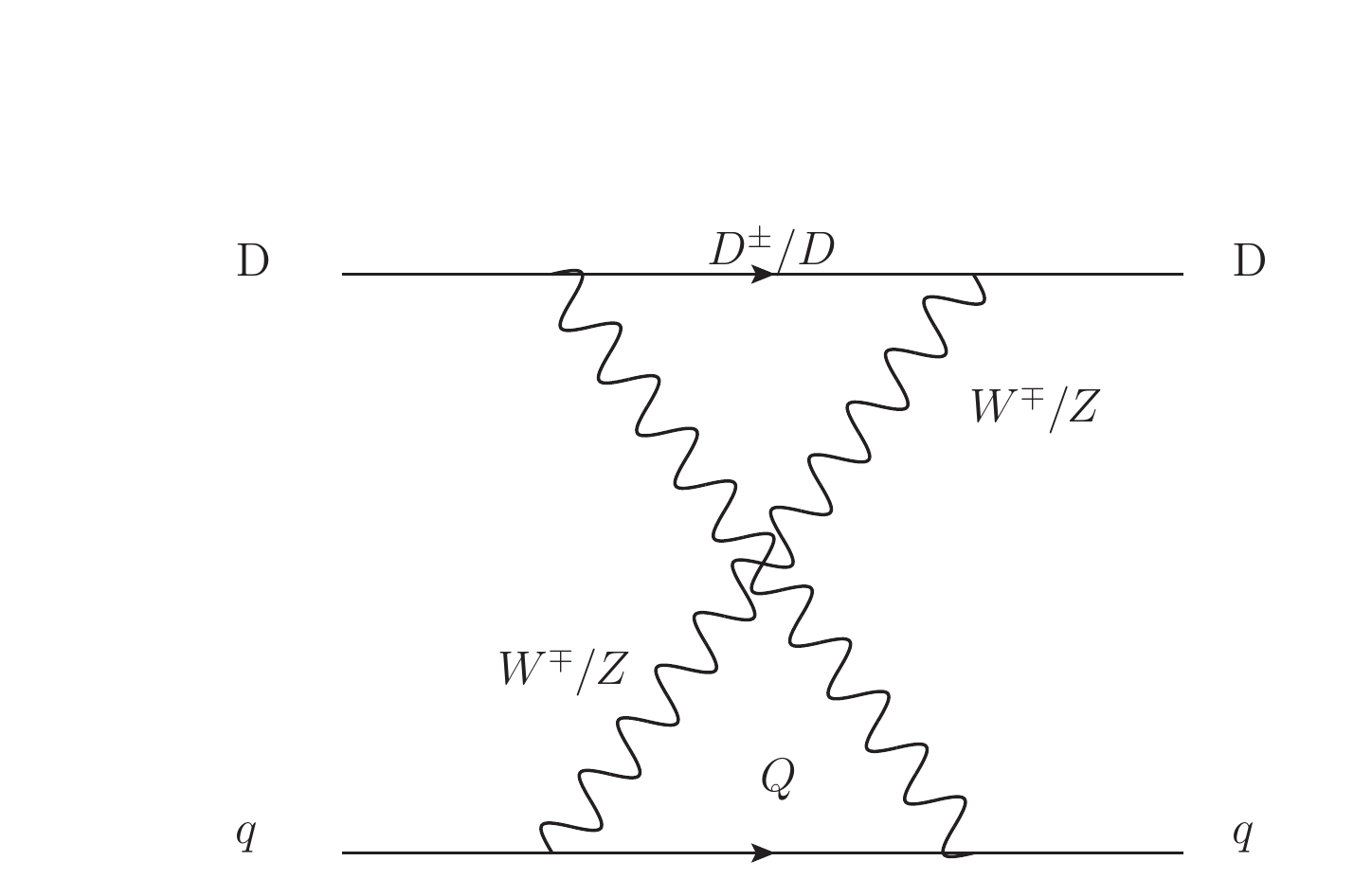}\\
\hspace*{0.2cm}\includegraphics[width=5cm]{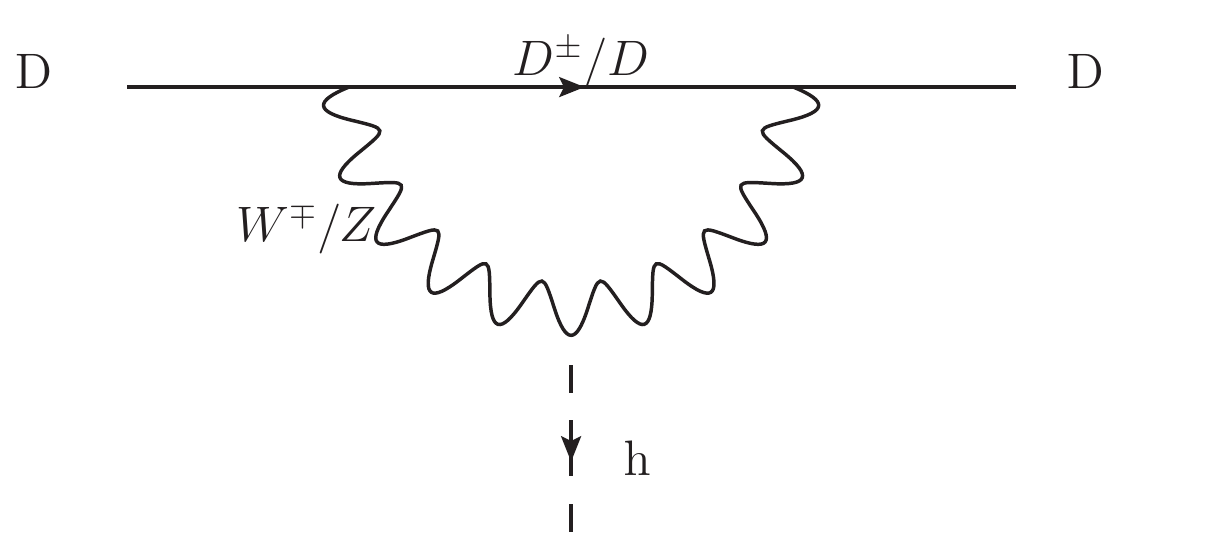}}
{\vskip -8cm\hspace*{5.0cm}\includegraphics[width=0.7\textwidth]{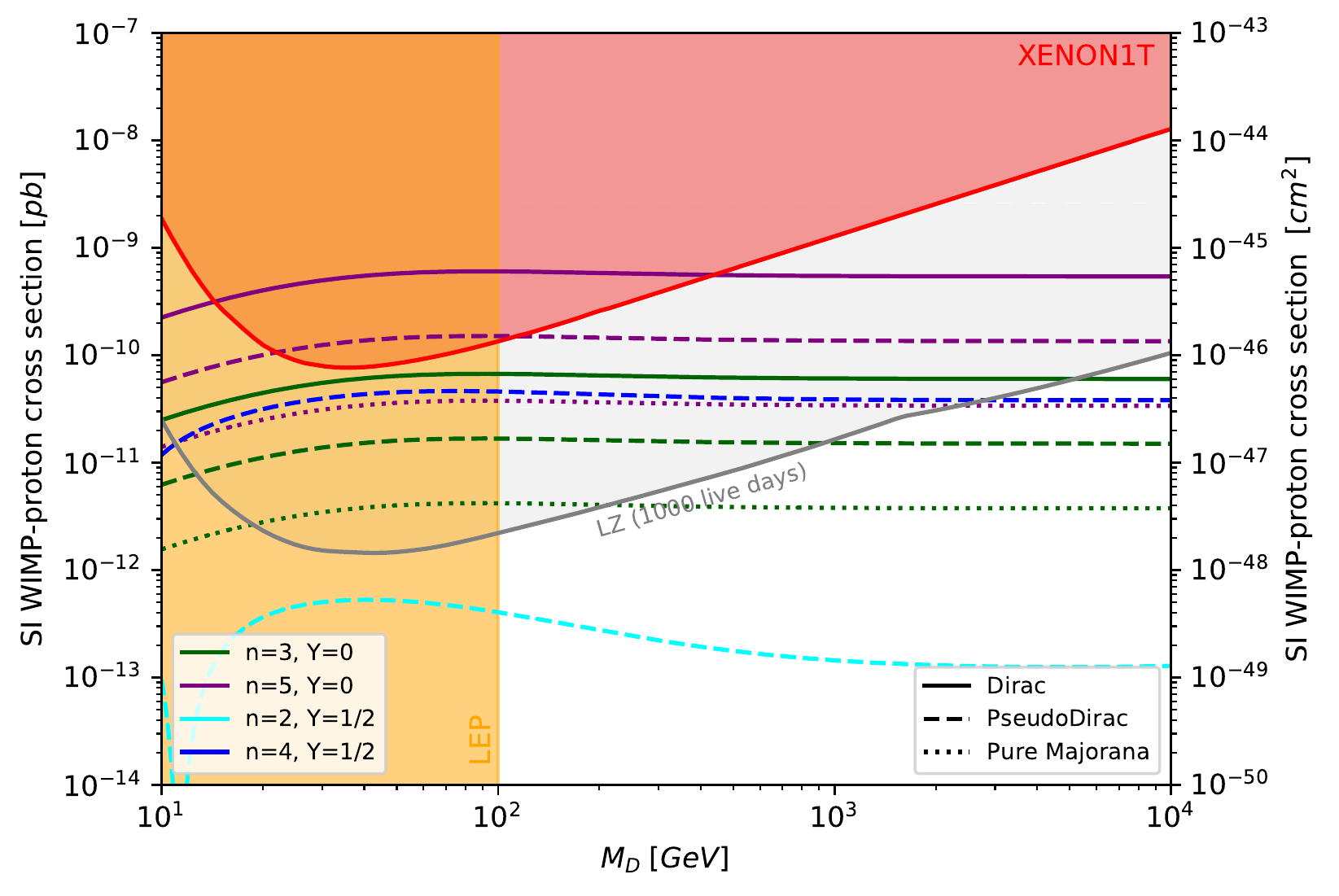}}
	\caption{Left: Loop diagrams for DM direct detection. Right: The spin-independent DM-proton cross section for a single fermion multiplet, for surviving cases $n\leq5$ for which the neutral component is the lightest 
	\label{fig:dmdd-loops}}
\end{figure}

One loop level direct detection in single multiplet DM models has been considered by several papers \cite{Cirelli:2005uq,Hisano:2011cs,Essig:2007az}. However, here we will extend these results to include the case of pure Dirac DM, and also consider the effect of the mass gap between DM and its partners that propagate inside the loops, diagrams for which are presented in Fig.~\ref{fig:dmdd-loops}(left).

One-loop induced direct detection rates and DM DD exclusion potential of the current  XENON 1T  experiment as well as future LZ experiments are presented in Fig.~\ref{fig:dmdd-loops}(right).
We have reproduced and extended results of paper~\cite{Hisano:2011cs} which noted cancellation between so-called twist-2 operator contribution and the rest of the 1-loop induced contributions which has been missed in~\cite{Cirelli:2005uq}. Due to this cancellation
XENON 1T  experiment is currently probing only (part of)  $(I=2, Y=5)$ Dirac DM model. The future LZ experiment, which will have two orders of magnitude higher sensitivity will be able to probe most of the models, except $(I=1,Y=0)$ Majorana DM model.


\section{Fermionic Dark Matter with one additional multiplet} \label{sec:mediator}

In the scenario of Dark Matter with one additional multiplet the  mediator multiplet can be either odd or even under the symmetry protecting the stability of  DM candidate, and its quantum numbers are limited (and defined) by the requirement of the renormalisability and gauge invariance of its interaction with DM multiplet.
We use different labels $F$/$\tilde{F}$ and $M$/$\tilde M$	for Dirac and Majorana fermion multiplets respectively
since they could define quite different models.
Here  we just list those models
which will be discussed in details in the follow up paper~\cite{MCDM-paper}:
\begin{itemize}
\item
	Even scalar mediator: $\tilde{F}^I_Y S^{I'}_{Y'}$ and $\tilde{M}^I_0 S^{I'}_0$.
The case of a scalar mediator that couples to the SM has been one of the first models considered in simplified scenarios (see e.g.~\cite{DiFranzo:2013vra,Buckley:2014fba,Baek:2015lna}), however it has been by now established that it is not simple nor minimal to achieve phenomenologically relevant models once the simplified case is included in a fully gauge-invariant model~\cite{Bauer:2017ota}. In particular, couplings to SM fermions are hard to obtain without breaking the EW symmetry, while couplings to gauge bosons only arise at  dim--5 operators level unless the scalar is allowed to develop a non-zero vacuum expectation value.
\item
Odd scalar mediator: $\tilde{F}^I_Y \tilde{S}^{I'}_{Y'}$ and $\tilde{M}^I_0 \tilde{S}^{I'}_{Y'}$.	
In this class of models, the DM fermion multiplet $\Psi$ couples to the odd scalar $\varphi$ and  to a SM fermion
via a Yukawa coupling: the quantum numbers of the scalar multiplet are, therefore, fixed by the properties of the  chosen  SM fermion.
As the SM fermions are  chiral, one can classify two cases, distinguished by their chirality, a SU(2)$_L$  doublet, $f_L$, or a singlet, $f_R$. 
\item
Even fermion mediator:
$\tilde{F}^I_Y F^{I'}_{Y'}$.
This case does not allow  renormalisable couplings between the mediator and the DM multiplet, however one could study the list of the respective operators  for completeness and because it leads to interesting new models of leptophilic DM.
\item
Odd fermion mediator: $\tilde{F}^I_Y \tilde{F}^{I'}_{Y'}$, $\tilde{M}^I_0 \tilde{F}^{I'}_{1/2}$ and $\tilde{F}^{I}_{1/2} \tilde{M}^{I'}_0$.
In the case of the odd fermionc mediators, the only renormalisable coupling is a Yukawa with the Higgs boson. In general, therefore, the DM state will be the lightest mass eigenstate from the neutral components of the two multiplets. Notable examples of this class of models come from SUSY, where the lightest neutralino can be a mixture of bino-Higgsino ($\tilde{M}_0^0 \tilde{F}^{1/2}_{1/2}$) or wino-Higgsino ($\tilde{M}^1_0 \tilde{F}^{1/2}_{1/2}$). Note that in our notation the first multiplet is the one that has the largest component in the DM physical state.
\item
Even vector mediators:
$\tilde{F}^I_Y V^{I'}_0$ and $\tilde{M}^I_0 V^{I'}_0$.
Vector mediators are very popular in the simplified model approach to DM phenomenology, mainly because they allow for ``gauge invariant'' couplings to vector current of SM fermions. Nevertheless, it is not a simple task to find a consistent, truly gauge invariant, renormalisable model containing vector mediator multiplets.
The easiest case  is the singlet, $V_0^0$, as it could arise from a broken gauged U(1) symmetry under which the SM fermions are charged. Though,  the consistent theory would require an anomaly-free U(1), thus either additional charged heavy states are added.
\item Odd vector mediators: $\tilde{F}^I_Y \tilde{V}^{I'}_{Y'}$.
In the case of odd vector mediators, the only allowed couplings must involve the DM multiplet and a SM fermion. 
Similarly to the case of even mediators, the above Lagrangian cannot be complete because of perturbative unitarity violation or the need to extend the gauge symmetries of the SM to generate $\tilde{V}$ as a gauge boson.
\end{itemize}


\section{Phenomenology of a new representative model:  $\tilde F^0_0S^0_0$(CP-odd)}
\label{sec:new-model-pheno}

let us  take a closer look at the  $\tilde F^0_0S^0_0$(CP-odd) model  with a
Dirac  fermion singlet ($\Psi \equiv \psi$) and a pseudo-scalar (CP-odd) singlet 
($\Phi\equiv a$) -- 
probably the simplest two component DM model  discussed in section 
\ref{sec:mediator}. 
The Lagrangian of the dark sector, to be added to the SM one, reads:
\begin{equation}
	\Delta \mathcal{L} = i \bar{\psi} \partial_\mu \gamma^\mu  \psi - m_\psi \bar{\psi}\psi + \frac{1}{2} (\partial_\mu a)^2 -  \frac{m_\Phi^2}{2} a^2 + i Y_\psi a \bar{\psi} \gamma^5 \psi  - \frac{\lambda_{aH}}{4} a^2 \phi_H^\dagger \phi_H - \frac{\lambda_a}{4}a^4\,,
\end{equation}
where $\phi_H$ is the SM Higgs doublet field,
$\phi$ is DM fermion and $a$ is the pseudo-scalar field. 
The model is described  by two masses:
$m_\psi$ and $m_a$ and three new couplings: the Yukawa coupling $Y_\psi$ connecting the 
scalar mediator $a$ to the fermion DM $\psi$, the $a$ self-interaction $\lambda_a$ 
and the quartic coupling to the Higgs ${\lambda_{aH}}$. The latter is the only coupling
connecting the new sector to the SM via a Higgs portal.  We recall that a linear coupling of $a$ to the Higgs
field is forbidden by CP.

Invariance under CP is preserved as long as $a$ does not develop a vacuum expectation value. We will be working in this region of the parameter space. As $\psi$ couples exclusively and bi-linearly to $a$, it is a stable fermionic DM candidate protected by a dark $U(1)$ global symmetry. The pseudo-scalar mediator $a$ can only decay into a pair of DM fermions. Hence, if $m_a < 2 m_\psi$, $a$ is said to be ``accidentally" stable and can contribute to the relic density as a second DM component: $a$ only couples bilinearly to the SM via the Higgs portal and only CP violation can allow for a linear coupling of $a$ to a SM operator. In this sense, it is the CP symmetry itself that prevents $a$ from decaying into SM states.

The interesting dynamics of this model, where $a$ is in touch with the SM via the Higgs portal coupling $\lambda_{aH}$, while $\psi$ only interacts with $a$,
leads to four distinct regimes of relevance for DM phenomenology,  summarised in table \ref{tab:PhenoRegimes}:

\begin{table}[htbp]
	\begin{tabular}{l|l|l|l}
		Scenario & $Y_\psi$                 & $\lambda_{aH}$               & DM thermal properties                                             \\ \hline
		A & $\mathcal{O}(10^{-3}-1)$ & $\mathcal{O}(10^{-3}-1)$ & $\psi$ and $a$ thermal with SM                            \\
		B & $<\mathcal{O}(10^{-8})$  & $\mathcal{O}(10^{-3}-1)$ & $\psi$ non-thermal, $a$ thermal with SM                   \\
		C & $\mathcal{O}(10^{-3}-1)$ & $<\mathcal{O}(10^{-8})$  & $\psi$ and $a$ thermal with each other, non-thermal to SM \\
		D & $<\mathcal{O}(10^{-8})$  & $<\mathcal{O}(10^{-8})$  & $\psi$ and $a$ non-thermal with each other and SM        
	\end{tabular}
	\caption{\label{tab:PhenoRegimes}
		Table  of distinct phenomenological DM scenarios possible in this model.}
\end{table}

\begin{itemize}
	\item  In scenario A, both fermion and pseudo-scalar can thermalise with the SM states. If $m_a\leq m_\psi$, then $a$ is stable and contributes to the relic abundance. Conversely, if $m_a>2m_\psi$, then it is unstable and merely acts as a mediator for the interactions of the fermionic DM to the SM. 
	
	\item In scenario B, the relic abundance of $\psi$ is driven by the freeze-in mechanism, while $a$ contributes as a thermal DM component for $m_a<2m_\psi$.  However, for $m_a>2m_\psi$, the smallness of $Y_\psi$ can lead to $a$ being metastable and decaying to (possibly warm) $\psi$. 
	
	\item In scenario C, both new particles can freeze-in via their couplings to the SM (the coupling of $\psi$ generated at loop level), before thermalisation between the two species. Depending on its mass, the pseudo-scalar $a$ can either remain  as a DM component, or decay promptly into the fermion DM $\psi$.
	
	\item In scenario D, both particles have very small couplings. While $a$ can freeze-in via its coupling to the Higgs portal, the coupling of the fermion is too small and would lead to a negligible direct production. Depending on its mass, $a$ can be the only significant DM candidate, or decay promptly to the fermion $\psi$ after being produced in the early universe. 
	
\end{itemize}

\noindent
Any other range of the couplings is excluded by DM over-production (or loss of perturbativity,).
Furthermore, in scenarios C and D, direct and indirect detection experiments, as well as colliders, would be unable to observe either of these new particles due to the feeble couplings. In contrast, in scenarios A and B, $a$ may be observable due to the sizeable Higgs portal coupling. In scenario A, the fermion may also be directly observables due to a loop-induced coupling to the Higgs.


The allowed regions of the parameter space should satisfy the relic density constraint from 
PLANCK~\cite{Adam:2015rua} ($\Omega_{\text{Planck}}h^{2}=0.1186\pm0.0020$, 
though we also allow under-abundant model points with $\Omega_h^2 < 0.12$, below PLANCK constraints), DM direct detection 
constraints from Xenon1T~\cite{Aprile:2017iyp,Aprile:2018dbl} (which are  dominant over the DM indirect detection constraints, as we have explicitly checked) and invisible  Higgs decay constraints from the 
LHC from ATLAS~\cite{ATLAS:2020kdi}  
(we use ${\mbox{Br} [H\to\mbox{invis}]<0.11}$).

\begin{figure}[htb]
	\includegraphics[width=0.5\textwidth]{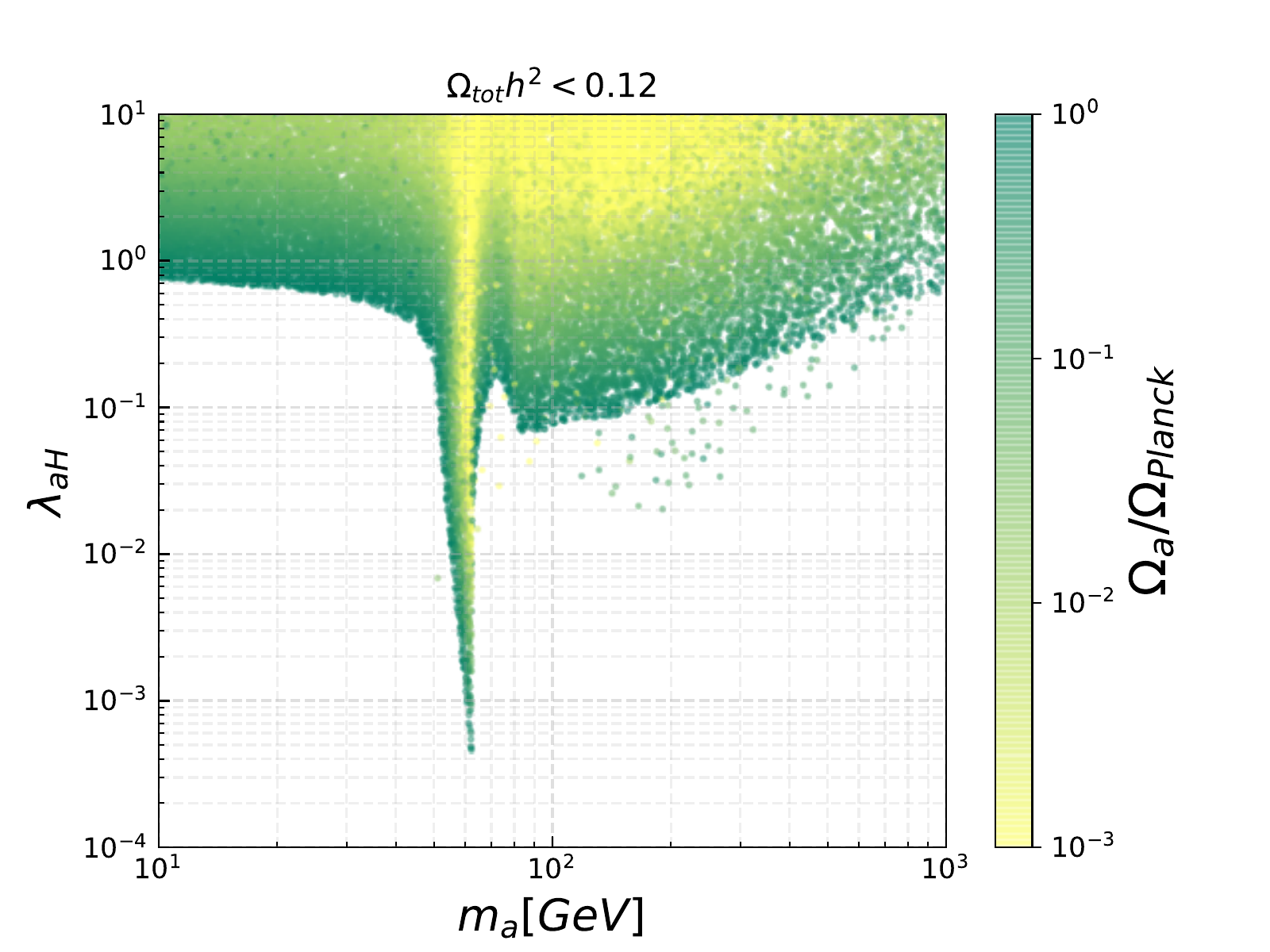}%
	\includegraphics[width=0.5\textwidth]{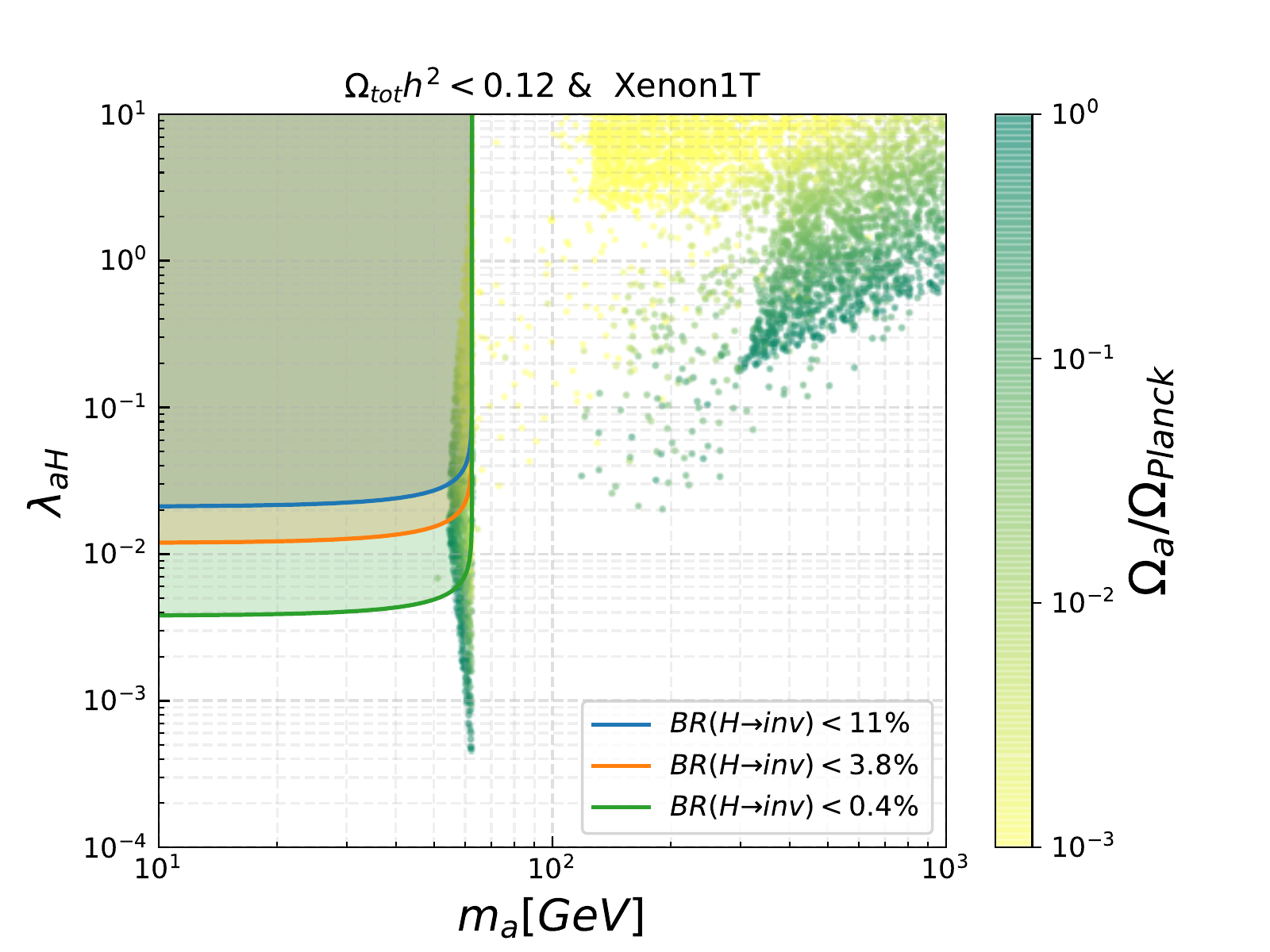}
	\caption{\label{fig:FDM-PS}
		2D projections of the allowed parameter space in ($m_a$, $\lambda_{aH}$) plane for $\tilde F^0_0S^0_0$ (CP-odd)  model (after constraints given at the top of each frame)
		with the colour map indicating the individual relative DM relic density.}
\end{figure}

As an example of our results for the scenario ``A", in Figure~\ref{fig:FDM-PS}, where we show the  2D projection of the  allowed parameter space in ($m_a$, $\lambda_{aH}$) plane after imposing the constraints listed in the top of the frame.
The colour map indicates the relic density normalised to the PLANCK value.

In  Figure~\ref{fig:FDM-PS} we show the projection of the allowed parameter space into the ($m_a$, $\lambda_{aH}$) plane, where
the colour map corresponds to values of $\Omega_a/\Omega_{Planck}$ with dark green marking model points that saturate the 
relic density with $a$ alone. 
The right panel~\ref{fig:FDM-PS}(a) present parameter space  surviving relic density constraint alone.
It clearly demonstrates the region of the resonant annihilation through the Higgs boson, 
$aa\to H$, which takes place for $m_a \simeq m_H/2$. Due to its efficiency, it 
allows the value of $\lambda_{aH}$ to go as low as $\simeq 4\times 10^{-4}$ 
while being consistent with the $\Omega_{\text{Planck}}$ constraint. Outside the
resonant region, values of $\lambda_{aH} \lesssim 10^{-1} \div 1$ are excluded by overclosure of the universe. Furthermore, in the right  panel~\ref{fig:FDM-PS}(b) we present the same 2D projection with points 
satisfying DM direct detection constraints from Xenon1T experiment (both on $a$ and on $\psi$).
The plot illustrates how Xenon1T excludes all points for $m_a \lesssim m_H$, except for a sliver
close to the Higgs resonance, where  $Haa$  coupling is small and/or relic density of  $a$ is low.
 We also show the LHC bound on the Higgs invisible decays
($\mbox{Br}[H\to\mbox{invis}]<0.11$), which excludes the Higgs resonant sliver for  $\lambda_{aH}\gtrsim 3\times 10^{-2}$,
as shown by the shaded region above the blue line. Future collider projections are considered as well, showing that
the exclusion on $\lambda_{aH}$ will improve by a factor of about 3 at the High Luminosity LHC run (HL-LHC) (projected bound of $\mbox{Br}[H\to\mbox{invis}] <  3.8\%$ \cite{Atlas:2019qfx}), as shown by the orange line. The International Linear Collider (ILC) running at $\sqrt{s}=250$~GeV and with an integrated luminosity of $1.15~\mbox{ab}^{-1}$ will be able to exclude $\lambda_{aH}\gtrsim 4\times 10^{-3}$, as indicated by the green line, corresponding to a projected exclusion of $\mbox{Br}[H\to\mbox{invis}]>0.4\%$ \cite{Asner:2013psa}. One should also note that even the ILC will not be able to fully exclude  the Higgs resonant region, where $\lambda_{aH}$ goes below the ILC sensitivity by one order of magnitude.


\section{Conclusions}

 We have performed a complete classification of MCDM models and  briefly discussed  features of some model classes. We found and studied new representative  model which had two DM candidates and provides viable DM candidate for both -- WIMP and FIMP scenarios.  We believe that this classification, and the MCDM approach, will create a solid framework for the consistent  complementary exploration of  DM 
at collider and non-collider experiments.

\section*{Acknowledgements}

AB is very grateful to organisers for their hospitality
and excellent organisation of the workshop.
Authors acknowledge the use of the IRIDIS High Performance Computing Facility, and associated support services
at the University of Southampton, in the completion of this work. 
AB acknowledges partial  support from the STFC grant ST/L000296/1 and Soton-FAPESP grant.

\bibliographystyle{moriond}
\bibliography{bib}
\end{document}